\documentclass[twocolumn,preprintnumbers,superscriptaddress,nofootinbib,aps,prd,floatfix,letterpaper]{revtex4}

\pdfoutput=1
\usepackage{amsmath, amssymb, bm, cancel, color, graphicx, enumerate, enumitem, appendix, youngtab, hyperref, bbold, verbatim, slashed}

\def\be{\begin{equation}}
\def\ee{\end{equation}}
\def\ba{\begin{eqnarray}}
\def\ea{\end{eqnarray}}

\def\p0{{\phantom{+}0}}

\def\uno{\mbox{1 \kern-.59em {\rm l}}}
\def\vincia{\texttt{VINCIA}}
\def\pythia{\texttt{PYTHIA}}
\def\herwig{\texttt{HERWIG}}
\def\sherpa{\texttt{SHERPA}}
\def\delphes{\texttt{Delphes}}
\def\fastjet{\texttt{FastJet}}

\definecolor{colorMD}{rgb}{1,0,0}

\definecolor{colorND}{rgb}{0,0,1}

\numberwithin{equation}{section}

\usepackage[english]{babel}

\begin{document}

\title{\Large{Parton Shower Uncertainties in Jet Substructure Analyses\\with Deep Neural Networks}}
\vspace{1cm}
\author{\small{\bf James Barnard}\thanks{\texttt{james.barnard@unimelb.edu}}}
\author{\small{\bf Edmund Noel Dawe}\thanks{\texttt{edmund.dawe@unimelb.edu.au}}}
\author{\small{\bf Matthew J. Dolan}\thanks{\texttt{dolan@unimelb.edu.au}}}
\author{\small{\bf Nina Rajcic}\thanks{\texttt{n.rajcic@student.unimelb.edu.au}}}
\affiliation{ARC Centre of Excellence for Particle Physics at the Terascale, School of Physics, University of Melbourne, 3010, Australia}

\begin{abstract}
  \noindent
Machine learning methods incorporating deep neural networks have been the
subject of recent proposals for new hadronic resonance taggers. These methods
require training on a dataset produced by an event generator where the true
class labels are known. However, this may bias the network towards learning
features associated with the approximations to QCD used in that generator which
are not present in real data. We therefore investigate the effects of
variations in the modelling of the parton shower on the performance of deep
neural network taggers using jet images from hadronic W-bosons at the LHC,
including detector-related effects. By investigating network performance on
samples from the Pythia, Herwig and Sherpa generators, we find differences of
up to fifty percent in background rejection for fixed signal efficiency. We
also introduce and study a method, which we dub zooming, for implementing
scale-invariance in neural network-based taggers. We find that this leads to an
improvement in performance across a wide range of jet transverse momenta. Our
results emphasise the importance of gaining a detailed understanding of what
aspects of jet physics these methods are exploiting.
\end{abstract}

\maketitle
\newpage

\section{Introduction}

The past decade has seen an explosion of interest in understanding and
exploiting the distribution of energy (substructure) within hadronic jets and
boosted resonances at the Large Hadron Collider
(LHC)~\cite{Abdesselam:2010pt,Altheimer:2012mn,Altheimer:2013yza,Adams:2015hiv}.
The study of jet substructure and the ability to identify (`tag') the hadronic
decay products of a wide variety of such resonances - the Higgs,  W and Z
bosons, top quarks, supersymmetric particles and other
Beyond-the-Standard-Model (BSM) states - is crucial in the analysis of both
Standard Model processes and in searches for BSM physics, which will only
become more important now that the LHC is running at high energy and with
future colliders on the horizon.

Since the foundational work in Ref.~\cite{Butterworth:2008iy} on studying jet
substructure in Higgs boson associated production, a multitude of taggers and
variables related to substructure have been
proposed~\cite{Butterworth:2008iy,Plehn:2009rk,Krohn:2009th,Ellis:2009me,Gallicchio:2010dq,Thaler:2010tr,Larkoski:2013eya,Larkoski:2014wba}.
These generally exploit our knowledge of QCD to construct functions which
effectively discriminate between signal and background. Some of these
techniques have already been applied to the problem of identifying boosted
massive vector bosons and top quarks by the ATLAS and CMS collaborations in Run
1 of the
LHC~\cite{Aad:2013oja,Aad:2013gja,Aad:2015owa,Chatrchyan:2012ku,CMS-PAS-JME-13-007,Khachatryan:2014vla,Khachatryan:2015oba}.

Another approach currently under development involves the application of
machine learning (ML) techniques to hadronic resonance tagging and searches for
new physics. The machine learning community has made large strides in problems
related to image recognition and computer learning, which may now also be
applied to particle physics. Signals produced by the LHC detectors may be
processed into pixelated jet images~\cite{Cogan:2014oua}, and machine learning
algorithms can be adapted to discriminate between a signal (such as $h\to b\bar
b$ decays or boosted hadronic top decays~\cite{Almeida:2015jua}) and background.
These algorithms have also been proposed as classifiers in neutrino
experiments~\cite{Racah:2016gnm,Aurisano:2016jvx}.

The use of machine learning, and neural networks in particular, has a long
history in particle physics and the idea of using neural networks for
quark-gluon
discrimination~\cite{Lonnblad:1990bi,Lonnblad:1990qp,Peterson:1993nk}, Higgs
tagging~\cite{Chiappetta:1993zv} and track identification~\cite{Denby:1987rk}
goes back over twenty-five years. However, the development of efficient deep
neural networks (DNNs) and the computing power associated with graphics
processing units (GPUs) means that image recognition technology has become
extremely powerful, driving the resurgence of interest in these techniques.

Recent work has seen the application of neural networks with two hidden layers
to hadronic top-quark tagging~\cite{Almeida:2015jua}, and deep convolutional
neural networks (known to have excellent performance in image classification)
to the problem of identification of hadronic W
decays~\cite{deOliveira:2015xxd}. These initial papers focussed on
demonstrating and understanding the network performance, and used truth-level
Monte Carlo (MC). The effects of pile-up and detector resolution were explored
in Ref.~\cite{Baldi:2016fql}, which showed that despite the loss of resolution
when these are taken into account the neural network is still somewhat superior
to traditional techniques. There has also been work on extending these methods
to jet flavour classification~\cite{Guest:2016iqz}.

The theoretical study of these techniques and their utility in high-energy
particle physics is still in its infancy, and there are a number of issues
still to be clarified in how deep learning methods may be applied at the LHC.
Some of these are related to the robustness of these techniques: how can we
guarantee that a network is learning about the physics differences between
signal and background, and not details particular to a specific MC event
generator? How robust are taggers based on these networks against detector
effects such as smearing and how do they degrade in the presence of pile-up? A
particular concern is that the network achieves a substantial fraction of its
discriminatory power from soft features in the spectrum which are modelled
phenomenologically rather than via perturbative QCD. This paper provides a
study of some of these issues.

We study the behaviour of neural networks over a number of different event
generators, and hence different parton showers and models for hadronisation.
For simplicity, we will usually just refer to these collectively as the parton
shower. 

We find that varying the parton shower leads to changes in the background
rejection efficiency of up to 50\%, depending on the shower model and selected
signal efficiency. We consider this to be large, and perhaps more than would be
expected from perturbative uncertainties from the parton shower. We believe
that caution is therefore required before these methods are applied on data,
and our results emphasise the necessity of understanding what features of the
jet images the neural networks are relying on to achieve their discriminatory
power. We also find changes in the factorisation and renormalisation scales
lead to negligible differences, while the addition of pile-up leads to an
overall degradation in network performance but not to a change in our
conclusions (in agreement with Ref.~\cite{Baldi:2016fql}).

There has also been interest recently in the
development of scale-invariant jet and substructure
taggers~\cite{Azatov:2013hya,Gouzevitch:2013qca,Dolen:2016kst,Schlaffer:2016axo},
and we discuss how similar ideas may be implemented in DNN-based taggers by
applying a $p_T$-dependent `zooming' factor on the jet images. The addition
of zooming leads to a slight improvement of around 10-20\% in the network
performance over a wide range of jet transverse momenta. 
While in this article we focus on discriminating between hadronically decaying
W bosons and QCD jets as a `standard candle', these methods should be
applicable to a wide variety of tagging and substructure issues. 

In Section~\ref{sec:arch} we outline the architecture and training of the
neural networks and in Section~\ref{sec:imagebuilding} discuss how we construct
jet images, and  present an idea of how to implement a scale-invariant tagger.
In Section~\ref{sec:event_gen} we show the variability in the DNN performance
across multiple event generators and parton shower models.

\section{Network Architecture, Training, and Performance Evaluation}
\label{sec:arch}

We follow Ref.~\cite{deOliveira:2015xxd} in our choice of network architecture,
who have already investigated the performance of a variety of different neural
networks. While we have investigated convolution networks, all results we
present here have been produced using the MaxOut~\cite{2013arXiv1302.4389G}
architecture. The network input consists of 625 units, equal to the number of
pixels ($25\times25$) present in each jet image. The input layer is followed by
two dense MaxOut layers consisting of 256 and 128 units each. The next two
layers are fully connected with 64 and 25 units and use a ReLU activation
function~\cite{ReLU}. The output layer consists of two nodes and a sigmoid
activation. Further discussion of network choices can be found
in Ref.~\cite{deOliveira:2015xxd}.

We used the Keras Deep Learning library~\cite{chollet2015keras} and the Adam
algorithm~\cite{KingmaB14} to train our networks on four NVIDIA Tesla K80 GPUs.
After selecting jet images within a window on the jet mass, $50 < m < 110$~GeV,
and transverse momentum, $200 < p_T < 500$~GeV, networks were trained with
approximately 3M signal and 3M background images where the signal and
background images have been weighted to produce flat $p_{T}$ distributions. A
portion (10\%) of the training images were set aside to evaluate a
cross-entropy loss function after each epoch and the network training
terminated after 100 epochs or after 10 epochs without an improvement in the
loss function. The Adam algorithm learning rate parameter was initially set to
0.001 and then reduced by 2\% after each epoch. We obtained reasonable
performance with a batch size of 100. We also implemented and tested a
cross-validated Bayesian optimisation procedure to determine optimal parameter
values but did not observe performance that was significantly better and so we
have left such investigations for future work. Further optimising the DNN
should anyway not affect our conclusions here as we probe the general
variability of a DNN with reasonable performance over different parton shower
models.

Finally, we evaluate the performance of a network by computing the inverse
background efficiency as a function of signal efficiency across a binned
likelihood ratio of the signal to background output of the network. This
variant of the standard receiver operating characteristic (ROC) curve better
displays differences in background rejection at low signal efficiency and can
be constructed from arbitrary jet observables or combinations of observables
through a (possibly multidimensional) binned likelihood ratio.

\section{Constructing a Jet Image}
\label{sec:imagebuilding}

This section provides a complete description of how we construct jet images,
from event generation to image output, along with the reasoning behind many of
our choices. Our process closely follows the one described in
Ref.~\cite{deOliveira:2015xxd} with the addition that we have also tested an
optional zooming step to reduce $p_{T}$-dependence.

We have developed a complete framework for jet image construction, network
training and performance evaluation. Low-level {\tt
Cython}~\cite{behnel2010cython} wrappers have been developed for
\pythia{}~\cite{Sjostrand:2014zea}, \delphes{}~\cite{deFavereau:2013fsa}, and
\fastjet{}~\cite{Cacciari:2011ma} that allow these tools to be connected in the
Python programming language, where particles, calorimeter towers, jets, and jet
images are stored as structured {\tt
NumPy}~\cite{Walt:2011:NAS:1957373.1957466} arrays and optionally written to
files on disk in the {\tt HDF5}~\cite{hdf5} format. This design provides the
potential to train networks on jet images generated `on-the-fly'. For the
studies presented here, however, we have created large {\tt HDF5} data sets of
jet images once that are then split and used for network training and
performance studies. Aside from the direct interface with \pythia{}, the
framework is able to study output from other event generators by reading
intermediate {\tt HepMC}~\cite{Dobbs:2001ck} files.

Following event generation, events are reconstructed with the \delphes{}
detector simulator while optionally overlaying pile-up events. Jets are then
reconstructed from calorimeter towers (referred to as jet constituents below)
with the anti-$k_t$ algorithm~\cite{Cacciari:2008gp} as implemented by
\fastjet{}~3.1.3. We have selected a jet clustering size of $R=1.0$ for all
studies presented here. For boosted $W$ bosons with two-body decays the
characteristic maximal separation of the subjets scales according to
\be
\Delta R=\frac{2m_W}{p_T^{\rm min}}
\ee
where $p_T^{\rm min}$ is the minimum transverse momentum of the jets to be
considered in the analysis. We have studied jets with transverse momenta above
200~GeV making $R=1.0$ a reasonable choice.

The highest $p_T$ jet is selected and subjets are formed in a jet
trimming~\cite{Krohn:2009th} stage, that also serves to lessen contributions
from soft radiation in the underlying event. Using the $k_t$ algorithm we
recluster the jet constituents into subjets with a fixed size of $r=0.3$ and
then discard all subjets with less than 5\% of the original jet momentum to
form a trimmed jet. All jet observables are computed with the trimmed jet.

The next stages are designed to remove spatial symmetries. First, all
constituents of the trimmed jet are translated in $\eta-\phi$ space to place
the leading subjet at the origin. We then define a grid of pixels with a
resolution of $0.1\times0.1$ in $\eta-\phi$ space and a jet image is
formed by taking the total transverse energy measured within each pixel
\be
E_{T,i}=\sum_{j}\frac{E_j}{\cosh{\eta_j}}
\ee
for all constituents $j$ in pixel $i$, with energy $E_j$ and original
pseudorapidity $\eta_j$. This image is rotated, either to put the subleading
subjet directly below the leading subjet or to align the principle component of
the jet image along the vertical axis if only one subjet is present. It is then
reflected, either to put the third-leading subjet on the right-hand side of the
image or to ensure that the total image intensity is highest on the right-hand
side if there are only two subjets. A cubic spline interpolation is used
wherever the transformed pixels do not align with the pixels of the original
image.

After the reflection stage above we are left with an image in which the leading
subjet is centred and the subleading subjet (if present) is directly below.
The separation between the two subjets is not constant, but varies linearly
with $2m/p_T$. By standardising this separation we can
potentially improve the DNN performance over a wide range of jet
$p_T$. The aim is to pick a scaling factor that enhances and standardises
features in signal images, i.e. those from boosted $W$ decays, without
artificially creating similar features in background, QCD images. This 
optional step is in addition to those detailed in Ref.~\cite{deOliveira:2015xxd}.

Denoting the physical separation between the two leading subjets as $\Delta
R_{\rm act}$, enlarging all jet images by a factor $R/\Delta R_{\rm act}$ for
some fixed $R$ gives a standardised jet image in which the separation (in
pixels) between the two leading subjets is fixed for all images. The downside
of this approach is that this is true for both signal and background images, so
an improvement in isolation of the subleading subjet is tempered by an
enhancement of signal-like features in background images. For this reason we
use the characteristic size assuming the $W$ mass, $s=2m_{W}/p_T$,
and enlarge all jet images by a scaling factor $\max(R/s, 1)$ where $R$ is the
original jet clustering size. The image enlargement is implemented by
interpolating at a finer resolution using a cubic spline. For signal images
$\Delta R_{\rm act}\approx s$ so this rescaling is very similar to, if a little
less effective than, using the actual separation of subjets to define the
scaling factor. For background images this scaling is not strongly correlated
with the subjet separation so the subleading subjet tends to be smeared out.

Jet images are then cropped to a fixed size of $25\times25$ pixels (whether
having been zoomed or not) and finally they are normalised to make the squared
sum of the pixel intensities evaluate to one. As discussed in
Ref.~\cite{deOliveira:2015xxd} this does not preserve the jet mass that can be
calculated from the original jet image but our zooming step destroys this
information anyway.

In summary, the full jet image construction and preprocessing steps are as
follows:
\begin{itemize}[leftmargin=*,itemsep=0em]
\item {\bf Jet Clustering and Trimming:} Reconstruct jets from calorimeter
  towers using the anti-$k_t$ algorithm with a jet size $R=1.0$ and select the
  leading jet. Trim the jet using the $k_t$ algorithm with a subjet size
  $r=0.3$.
\item {\bf Translation:} Translate all jet constituents in $\eta-\phi$ space to
  put the leading subjet at the origin.
\item {\bf Pixelisation:} Pixelise the transverse energy of the jet using
  pixels of size $(0.1, 0.1)$ in $\eta-\phi$ space. This produces a jet image.
\item {\bf Rotation:} Rotate the jet image to put the subleading subjet
  directly below the leading subjet. If no subjets are present rotate to align
  the principle component of the jet image along the vertical axis.
\item {\bf Reflection:} Reflect the jet image in the horizontal direction to
  put the third-leading subjet on the right-hand side. If there are only two
  subjets reflect to ensure that the summed image intensity is highest on the
  right-hand side.
\item {\bf Zooming:} Optionally zoom the jet image by a factor that reduces
  dependence on the jet momentum.
\item {\bf Cropping and Normalisation:}
  Crop the jet image at $25\times25$ pixels and normalise pixel intensities to
  make the sum of their squares equal to one.
\end{itemize}

\begin{figure}
  \centering
  \includegraphics[width=\linewidth]{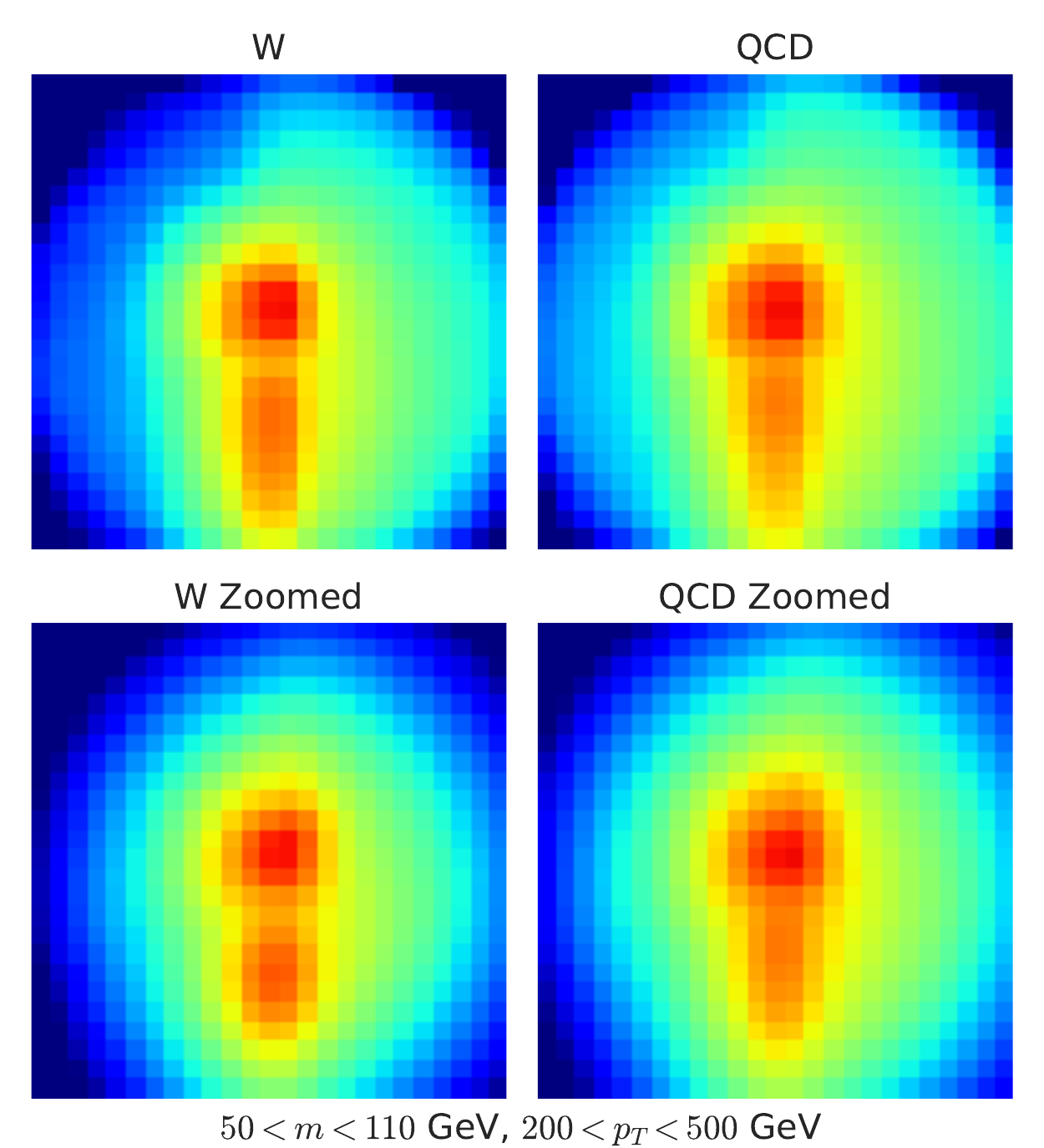}
  \caption{We show the average jet images obtained for hadronic W bosons and
    QCD as modelled by the \pythia{} default shower. The images on the top have
    been preprocessed in the standard way, while those on the bottom have also
    undergone the zooming procedure outlined in
    Section~\ref{sec:imagebuilding}. The axes are left unlabelled since they do not
    correspond to the physical $\eta$ and $\phi$ dimensions following image
    rotations, reflections, and zooming. Pixels are coloured according
    to higher (red) and lower (blue) average normalised pixel intensities.}
  \label{fig:zoomed}
\end{figure}

\begin{figure}
  \centering
  \includegraphics[width=\linewidth]{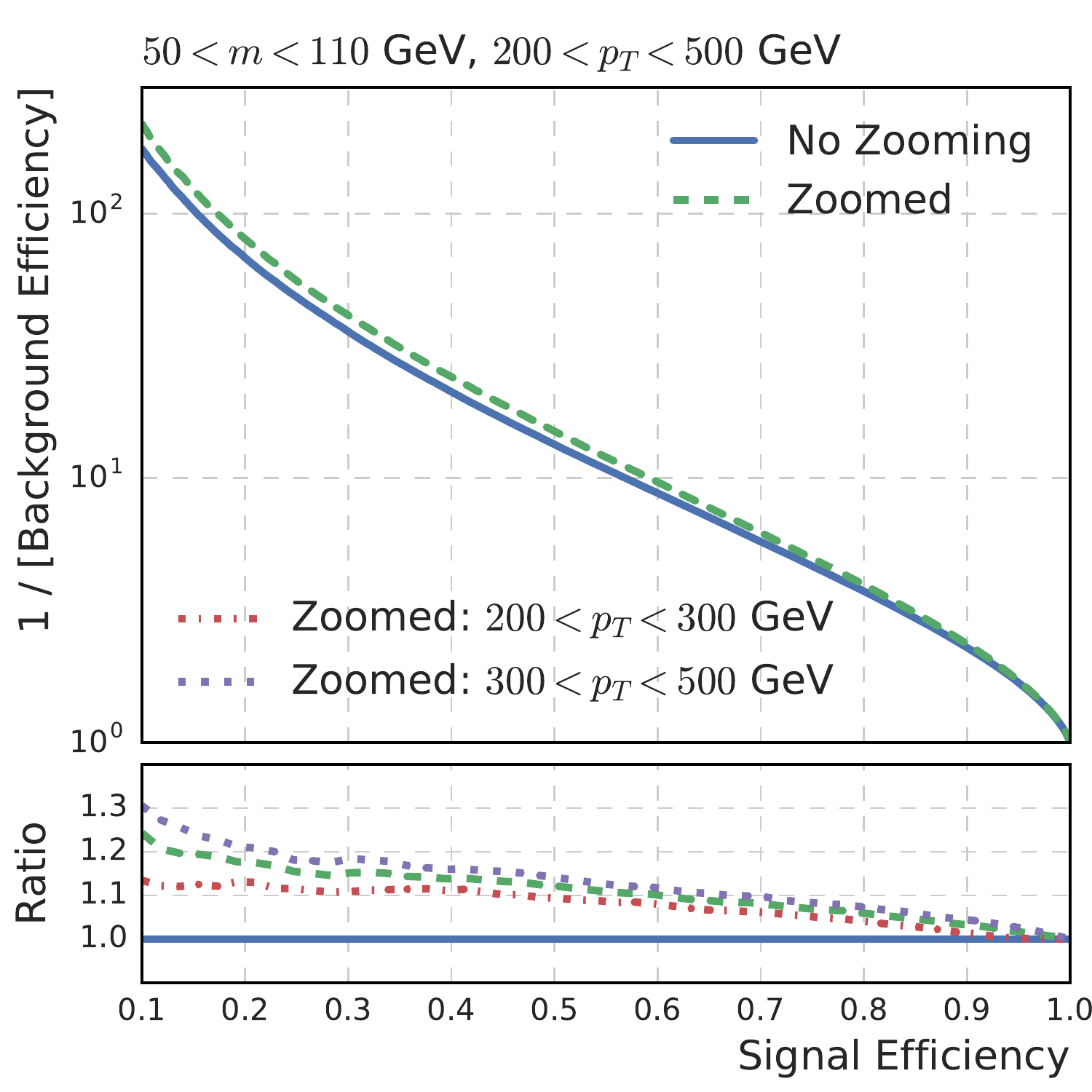}
  \caption{The ROC curves for the zoomed (solid blue) and unzoomed (dashed
    green) jet images for the \pythia{} default shower. The lower panel shows
    the ratio of the zoomed to unzoomed efficiencies, also showing the
    efficiency sliced in bins from $200 < p_T < 300$~GeV (dot-dashed red) and
    $300 < p_T < 500$~GeV (short dashed blue).\label{fig:zoom_compare}}
\end{figure}

In Fig.~\ref{fig:zoomed} we show the average jet images for boosted W and QCD
jets in the range $200 < p_T < 500$~GeV for the default \pythia{} shower
using the standard preprocessing in the top panels, and using the zooming
procedure in the bottom panels. For the W-jets we note the zooming procedure
results in a more regular and compact average shower shape, and that the second
(lower $p_T$) subjet becomes better spatially defined as expected. While the
average image of the QCD jets becomes more compact, the subjets
remain somewhat smeared compared with the W-jets. Since the subjets do not
originate in the decay of a heavy resonance and hence are not associated with a
specific mass scale this is not a surprise. 

An obvious conceptual advantage of using the zooming technique is that it makes
the construction of scale-invariant taggers easier. Scale invariant
searches~\cite{Gouzevitch:2013qca,Azatov:2013hya,Dolen:2016kst,Schlaffer:2016axo}
which are able to interpolate between the boosted and resolved parts of phase
space have the advantage of being applicable over a broad range of masses and
kinematics, allowing a single search or analysis to be effective where
previously more than one may have been necessary. 

We show in Fig.~\ref{fig:zoom_compare} the ROC curves for two different neural
networks: the first (the solid blue line) was trained without zooming, while
the second (the green dashed line) used zooming. Both networks were trained and
tested on samples of jet images in the mass window  $50 < m < 110$~GeV and a
large $p_T$ range, $200 < p_T < 500$~GeV. As predicted, the zoomed network
outperforms the unzoomed one, particularly at low signal efficiency, where the
background rejection rises by around 20\%. We obtain similar results when we do
not restrict the sample of jet images within a mass window. We find that the
zooming has the greatest effect at high $p_T$. For less boosted $W$ decays the
enhancement in background rejection is around 10\% which rises to just over
20\% for $300 < p_T < 500$~GeV.

\section{Event Generator Dependency}
\label{sec:event_gen}

\begin{figure}
  \centering
  \includegraphics[width=0.49\linewidth]{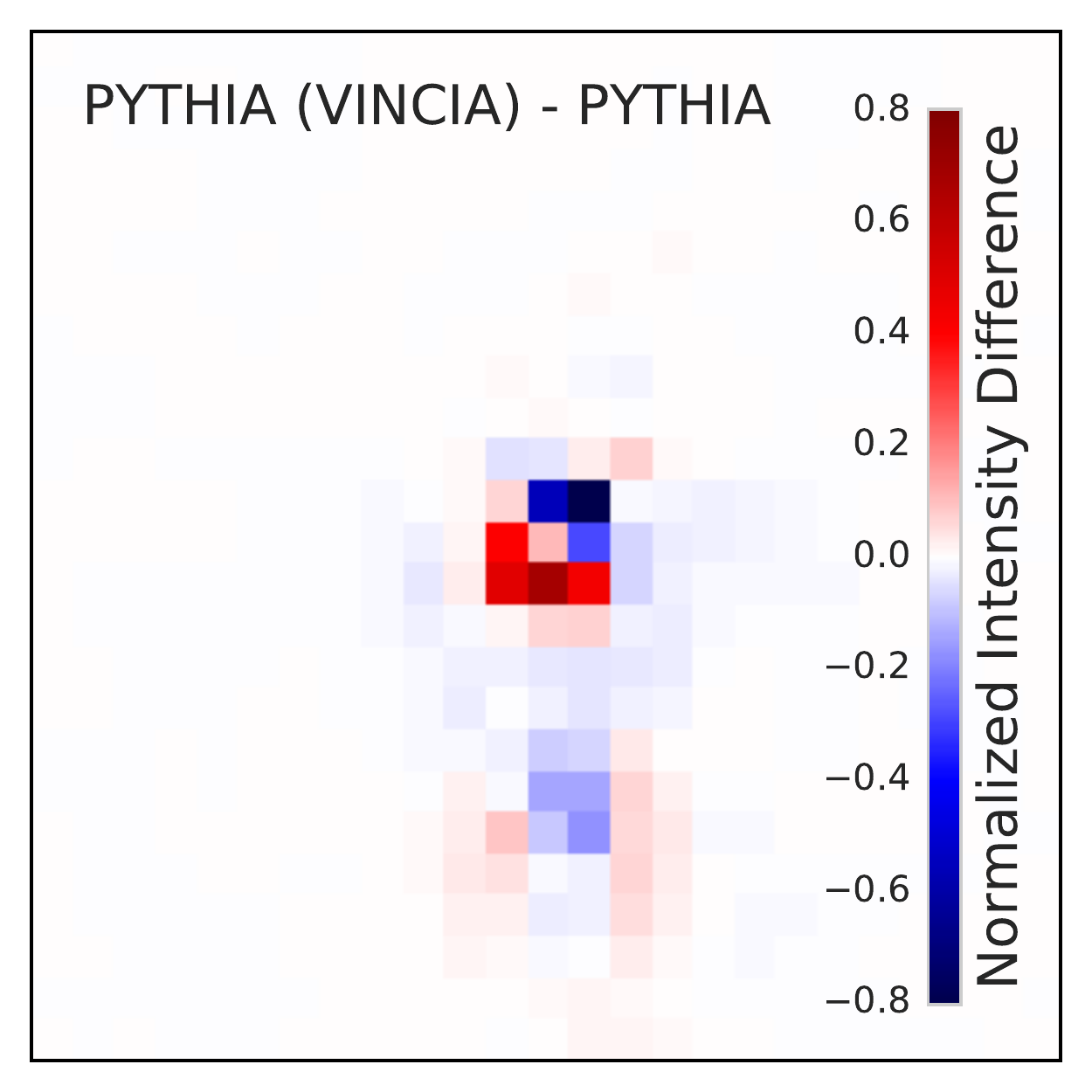}
  \includegraphics[width=0.49\linewidth]{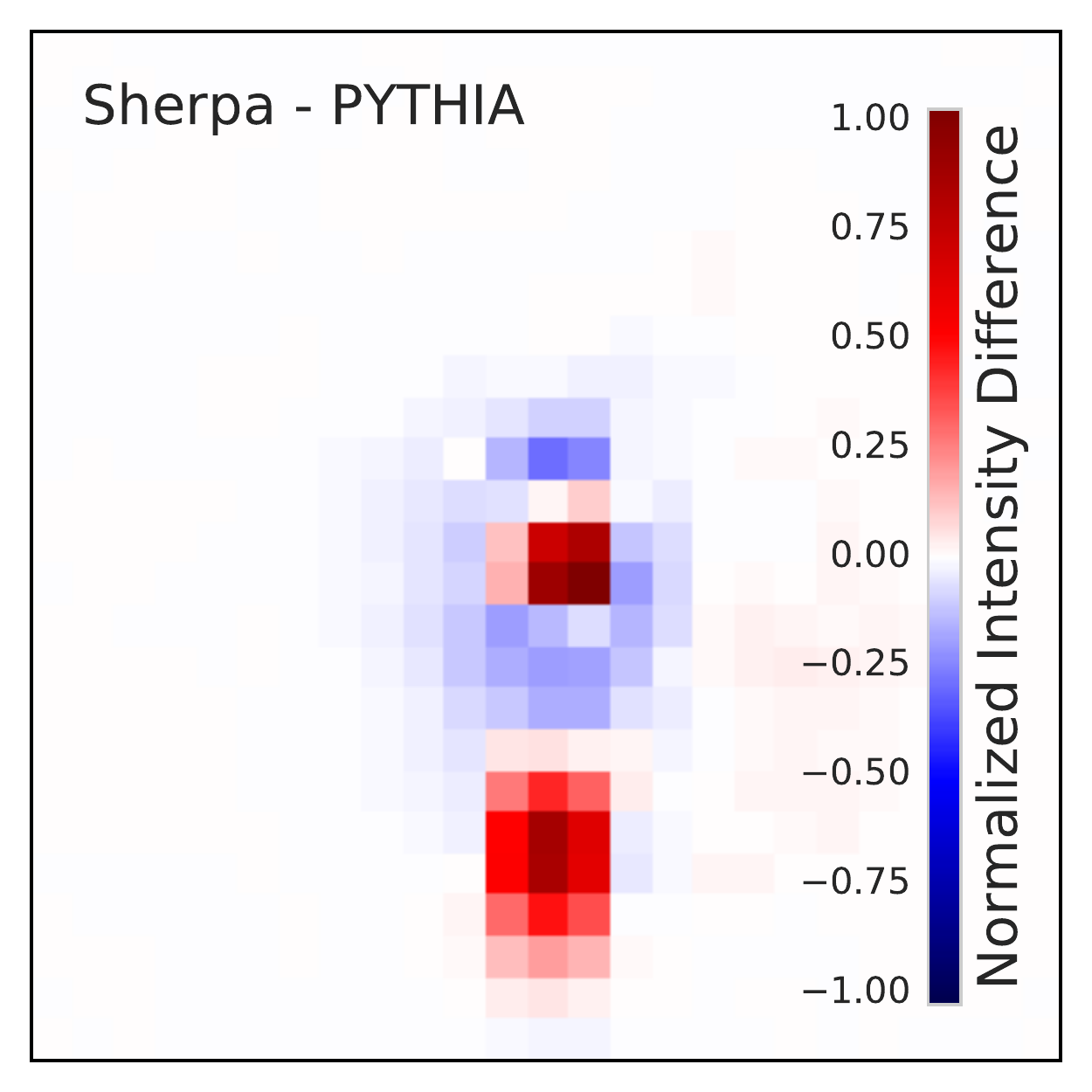}\\
  \includegraphics[width=0.49\linewidth]{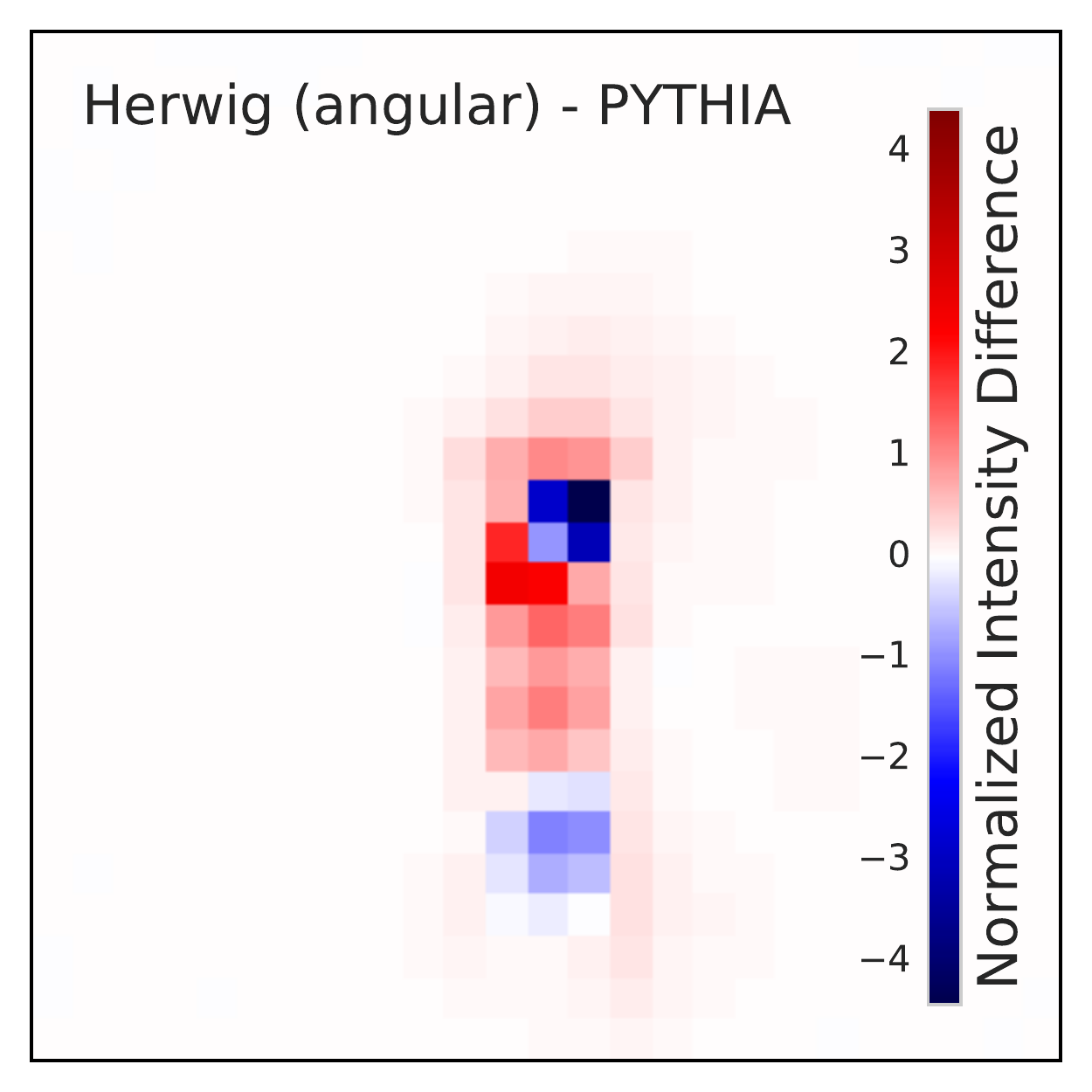}
  \includegraphics[width=0.49\linewidth]{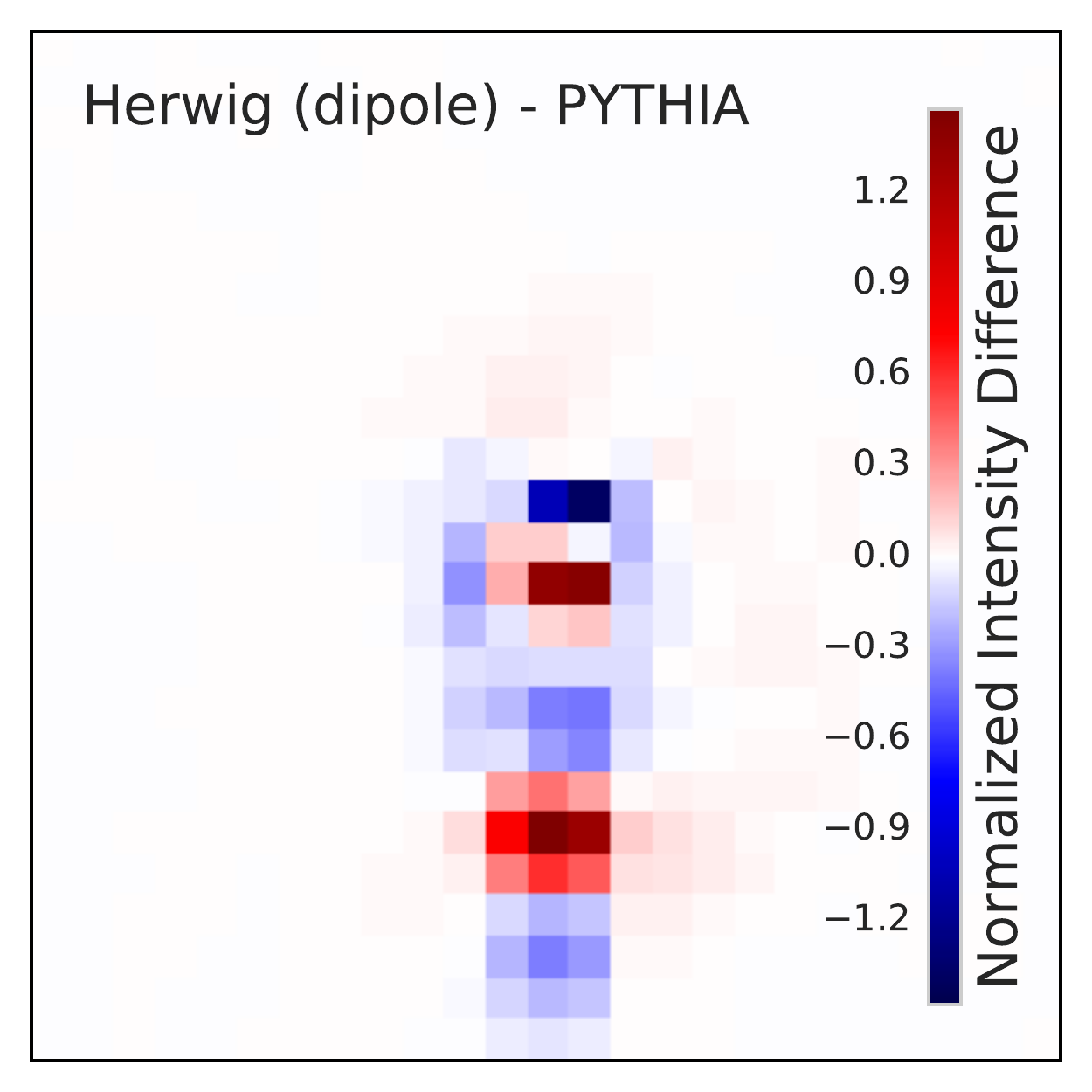}
    \caption{This figure shows the W-jet image differences between the default \pythia{}
  shower and the alternate \vincia{} shower in \pythia{} (top left), the default \sherpa{}
  shower (top right), the default \herwig{} angular shower (bottom left) and the
  \herwig{} dipole shower (bottom right). The plots have been individually normalised.\label{fig:shower_compare}}
\end{figure}

The networks require supervised training prior to being applied on unlabelled
data. Since it is difficult to isolate regions of very high signal purity,
training on simulated data is necessary before application to real LHC events.
However, all MC event generators and parton showers are only approximations to
the full Standard Model. Understanding what features of QCD a DNN is learning
about, and whether it is learning event generator-dependent approximations is
thus an important question. Furthermore, there are features of real-world QCD
such as colour reconnection which, while modelled in the parton shower, are in
reality poorly understood. We will not attempt to quantify those effects in
this work.
\begin{figure*}[t!]
  \centering
  \includegraphics[width=0.49\linewidth]{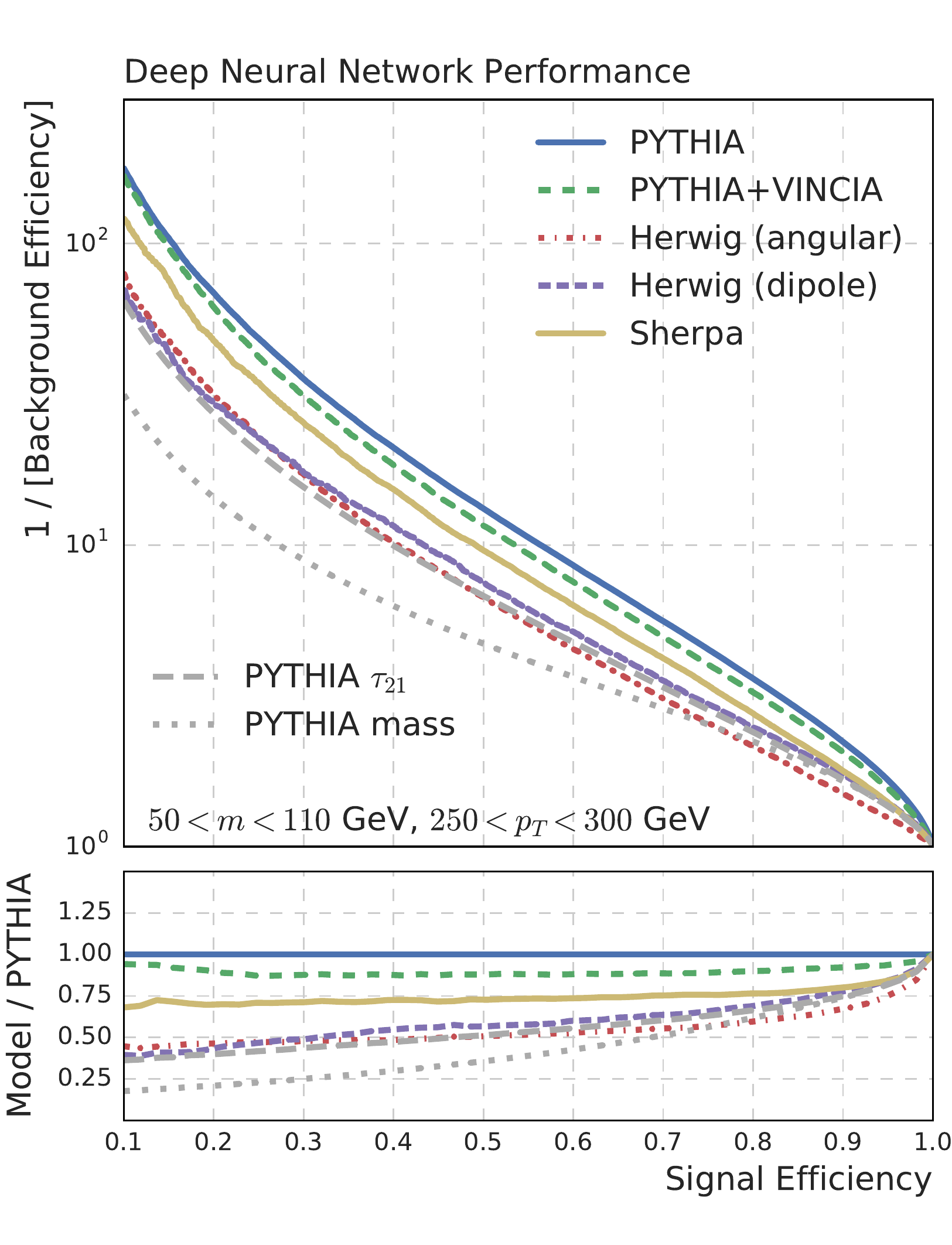}
  \includegraphics[width=0.49\linewidth]{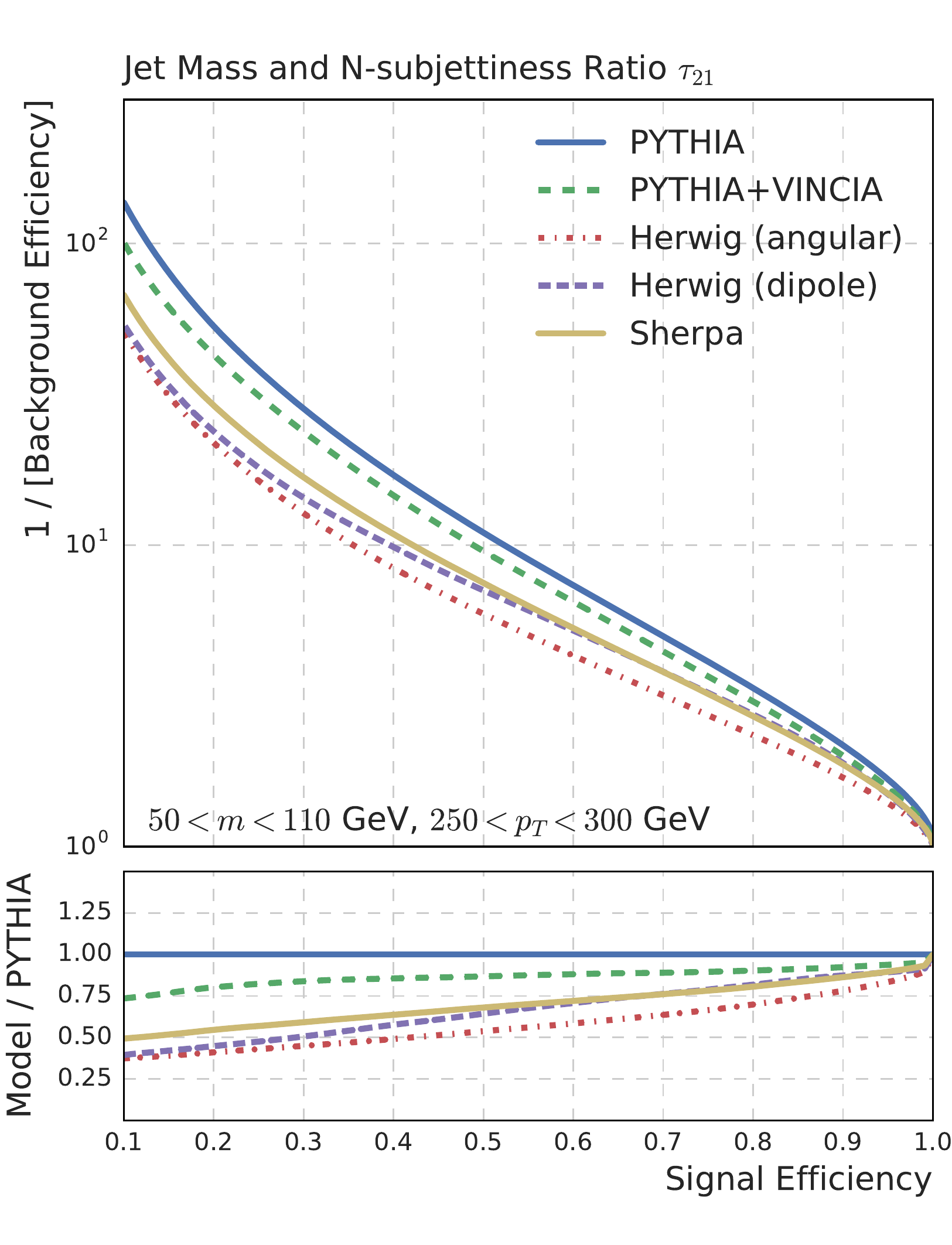}
  \caption{This figure shows the ROC curves of the \pythia{} (solid blue),
  \vincia{} (dashed green), \herwig{} angular (red dash-dot) and dipole (dashed
  purple), and \sherpa{} (solid gold) showers for the DNN output (left) and the
  combination of the jet mass and $n$-subjettiness ratio $\tau_{21}$ through a
  two-dimensional binned likelihood ratio (right). The lower panels show the
  ratio of the ROCs with the default \pythia{} shower. All ROC curves are computed
  using jet images within a window on the jet mass, $50 < m < 110$~GeV, and
  transverse momentum, $250 < p_T < 300$~GeV.}
\label{fig:roc_compare}
\end{figure*}

To gain an understanding of the systematic uncertainties in using networks
trained on simulated data, we study the behaviour of networks across a variety
of different generators and parton showers which all provide an adequate
description of current LHC data. We assume that given a number of different ROC
curves derived from different generators and parton showers, the envelope of
these curves provides an approximate uncertainty band associated with training
the network on simulated, rather than real, data.

Recently, Ref.~\cite{Bellm:2016rhh} has studied parton shower uncertainties in
\herwig{}~7. They divide the uncertainties into a number of classes: numerical,
parametric, algorithmic, perturbative and phenomenological. Numerical
uncertainties can be decreased by increasing the number of events, while
parametric uncertainties are those external to the MC generator: masses,
couplings, PDFs and so forth. The focus of our work in this section is on
algorithmic uncertainties, those due to different choices of parton shower
algorithm. The authors of Ref.~\cite{Bellm:2016rhh} focus on perturbative and
phenomenological uncertainties, which are from truncation of expansion series
and parameters deriving from non-perturbative models. Our work is more in the
spirit of the `Towards parton shower variations' contribution to the 2015 SM
Les Houches Proceedings~\cite{Badger:2016bpw}. Previous studies also exist
within the \herwig{} framework on the implications of MC uncertainties on jet
substructure in the context of Higgs searches~\cite{Richardson:2012bn}.

We generate background and signal events with three of the most widely used MC
generators: \pythia{}~8.219~\cite{Sjostrand:2014zea},
\sherpa{}~2.0~\cite{Gleisberg:2008fv,Gleisberg:2008ta} and
\herwig{}~7.0~\cite{Bahr:2008pv,Bellm:2015jjp}. For \pythia{}~8 we study both
the default shower and the \vincia{}
shower~\cite{Fischer:2016vfv,Ritzmann:2012ca}, and for \herwig{} we include
both the default (angular ordered) and dipole
showers~\cite{Platzer:2009jq,Platzer:2011bc}, giving us five different parton
shower models to study.

The default \herwig{} shower (known as QTilde) is based on $1\to 2$ splittings
using the DGLAP equations, with an angular ordering
criterion~\cite{Gieseke:2003rz}. The \sherpa{} shower is based on a
Catani-Seymour dipole formalism~\cite{Schumann:2007mg}. In this case one
particle of the dipole is the emitter which undergoes the splitting, while the
other is a spectator which compensates for the recoil from the splitting and
ensures that all particles remain on their mass-shells throughout the shower,
leading to easier integration with matching and merging techniques. The default
shower in \pythia{}~8 is also a dipole style shower~\cite{Sjostrand:2004ef},
ordered in transverse momentum.

While parton showers have traditionally been based upon partonic DGLAP
splitting functions, another possibility is to consider colour-connected parton
pairs which undergo $2\rightarrow 3$ branchings (note that this is distinct
from Catani-Seymour dipoles used in \sherpa{}, where one parton is still an
emitter, and the other recoils). In these so-called antenna showers, the
2-parton antenna is described with a single radiation kernel. This has the
advantage, for instance, of explicitly including both the soft and collinear
limits. We use the recently released
\vincia{}~\cite{Fischer:2016vfv,Ritzmann:2012ca} plug-in for \pythia{}~8
as a representative antenna shower.

These event generators also provide different treatments of the soft
radiation from the underlying event which accompanies each hard partonic
scattering. They also possess different
implementations of the parton-to-hadron fragmentation process being based
either around cluster fragmentation ideas (\herwig{} and \sherpa{}) or the Lund
string model (\pythia{}), giving us a wide range of QCD-related effects to
probe. To incorporate detector effects such as smearing we pass all events
through the \delphes{}~3 detector simulator~\cite{deFavereau:2013fsa}. In the
studies presented here, our baseline shower is \pythia{}~8 with its default
settings.

We construct average jet images for all five different generators and showers
under investigation, and then subtract the default \pythia{} average jet image
in order to see the differences in the average radiation patterns. The results
are shown in Fig.~\ref{fig:shower_compare} for the W-jet signal. We have
normalised the intensity differences of the pixels so that red indicates a
region of excess and blue a deficit relative to the \pythia{} default. While the
\vincia{} is roughly similar to the
\pythia{} default, the \sherpa{} and \herwig{} dipole showers  exhibit more intense radiation in the resolved subjets and a
substantial deficit in the region between the subjets. The \herwig{} angular shower shows the opposite, with less radiation in the subjet cores and more diffuse radiatioon. QCD radiation exhibits
similar features.

Next we show ROC curves for the different showers in
Fig.~\ref{fig:roc_compare}. We used the same network discussed in
Section~\ref{sec:imagebuilding} trained on the default \pythia{} shower
(without zooming), and then used events from the other generators and parton
showers as input, \textit{e.g.} we ask a neural network trained on the
\pythia{} shower to discriminate between QCD and W-jets from \sherpa{}.

We do not extend the ROC curves down to zero signal efficiency since they are
more statistically limited there. The \pythia{} ROC is higher than all other shower efficiency curves.
While both the \sherpa{} and \herwig{} dipole images exhibit
superficial similarities in Fig.~\ref{fig:shower_compare}, the network is
better at discriminating the \sherpa{} events. At a fixed low signal efficiency the
\herwig{} angular and dipole showers have the lowest background rejection, smaller than that
obtained using the \pythia{} default by a factor of two. The \vincia{} and
\sherpa{} showers have a slightly lower rejection rate than the \pythia{} one.
For signal efficiency of 50\% the uncertainty from changing the event generator
is around 40\%. 

For large background rejection rate we note that the network trained on the
\pythia{} events has a lower efficiency for selecting signal events generated
from the other showers, \textit{i.e.} it is maximally efficient for the shower
it was trained on. This may be due to the network learning some features
associated specifically with the \pythia{} shower and thus performing well
on \pythia{}-like events.

We also show in Fig.~\ref{fig:roc_compare} the ROC curves we obtain for the
trimmed jet mass and the $n$-subjettiness ratio $\tau_{21}\equiv \tau_{2}/
\tau_{1}$~\cite{Thaler:2010tr} which is often used as a discriminating variable
in studies of jet substructure~\cite{ATL-PHYS-PUB-2014-004}. We see that the
neural network consistently outperforms these variables (in agreement with the
conclusions already reached in Ref.~\cite{deOliveira:2015xxd}). This result stands
independent of the uncertainty induced by the choice of event generator, although the results for the \herwig{} showers are close to being degenerate with it.

In the right panel of Fig.~\ref{fig:roc_compare} we show the ROC curves we
obtain from the combined jet mass and $\tau_{21}$ observables for the different
parton showers. We see that the parton shower uncertainties in this case are
very similar to those obtained from the jet images. The uncertainties from the
varying the parton shower for the jet images are thus of similar size to those
associated other more common variables, such as those found in theoretical studies of the D2 tagger~\cite{Larkoski:2014gra,Larkoski:2015kga} and those found by the ATLAS Collaboration in~\cite{Aad:2015rpa} in searches for boosted W-bosons. 

Another possible independent source of uncertainty is the dependence of the
shower profile on the common renormalisation and factorisation scales, $\mu_R$
and $\mu_F$ respectively which for our purposes are set to be $\mu=p_T$. As is
standard, we vary the scale $\mu$ upwards and downwards by a factor of two from
its default value of $\mu=p_{T,W}$. We find that the changes in ROC curves due
to this were negligible.

\section{Conclusions}
\label{sec:conc}

The use of deep neural networks to construct classifiers for hadronic
substructure using jet images is an exciting proposal. However, it is important
to quantify the dependence on the training dataset, and whether the network is
learning the approximations inherent in the MC generator. We trained a network
on the default parton shower from the \pythia{} generator and studied its
performance on events from \herwig{} and \sherpa{}. We found that the network performed better on test events also from the default \pythia{}
shower, indicating that the network may be learning some \pythia{}-specific
features. The change in performance through using different parton showers
could be up to a factor of two in background rejection. Our results thus
indicate that care is required to avoid over-interpretation of small changes in
ROC curves, given the parton shower uncertainties. These uncertainties are
relatively large, and further study is required to ascertain whether the
network performance is truly being driven by features in the parton shower
(which are under control in perturbative QCD) or by softer physics such as
hadronisation modelling (which is not). Either way, our results demonstrate
that caution is required in the application of machine learning techniques on
simulated data. We intend to return to this issue in the near future.

There are many opportunities for further work in this area. One way to achieve
event generator independency is by avoiding the use of training data through
using data-driven unsupervised learning algorithms. Will these prove as
powerful as the supervised techniques using DNNs proposed thus far? It would
also be desirable to incorporate more than just calorimeter information into
jet images, in a similar vein to recent work on heavy flavour
tagging~\cite{Guest:2016iqz}. We note that tracking information has been
proposed in the context of substructure as being particularly important at high
energies~\cite{Spannowsky:2015eba}. Since jet images can be both large and
sparse, new algorithms may be required to render this
feasible~\cite{DBLP:journals/corr/MnihHGK14}. In an ideal world, it would be
possible to use information from the whole detector to classify events into
signal or background, as in event deconstruction~\cite{Soper:2014rya}.

The main outcome of our study is to emphasise the importance of what the neural
networks are learning and how they use it to discriminate between signal and
background. This lesson is true for any application of machine learning in
particle physics, from widely used techniques such as boosted decision trees to
newer methods like the deep neural network we have studied. However, provided
cautious and detailed studies of the uncertainties involved lead to methods to
constrain them in the analysis of real data, these methods may prove to be
powerful and reliable analysis tools for future searches and measurements.

\section*{Acknowledgements}
We thank Peter Skands, Michael Spannowsky and Phillip Urquijo for helpful
discussions and comments. This work was supported in part by the Australian
Research Council. We are grateful for the computational support provided by our
colleagues in CoEPP and Research Compute Services at the University of
Melbourne for generously providing access to the GPUs. We thank the Aspen
Center for Physics, which is supported by National Science Foundation grant
PHY-1066293, the Munich Institute for Astro and Particle Physics and the Mainz
Institute of Theoretical Physics, where part of this work was completed, for
their hospitality and support.

\bibliographystyle{JHEP}
\bibliography{deepjets}

\end{document}